\documentclass{aastex631}

\usepackage{color,xspace}

\shorttitle{Megaconstellation visibility}
\shortauthors{Lawler, Boley \& Rein}

\graphicspath{{./}{figures/}}

\begin{document}

\title{Visibility Predictions for Near-Future Satellite Megaconstellations: Latitudes near 50$^{\circ}$ will Experience the Worst Light Pollution}

\correspondingauthor{S.~M. Lawler}
\email{samantha.lawler@uregina.ca}

\author[0000-0001-5368-386X]{Samantha~M. Lawler}
\affiliation{Campion College and the Department of Physics, University of Regina}
\author[0000-0002-0574-4418]{Aaron C.~Boley}
\affiliation{Department of Physics and Astronomy, University of British Columbia}
\author[0000-0003-1927-731X]{Hanno Rein}
\affiliation{Department of Physical and Environmental Sciences, University of Toronto at Scarborough}

\begin{abstract}

Megaconstellations of thousands to tens of thousands of artificial satellites (satcons) are rapidly being developed and launched.
These satcons will have negative consequences for observational astronomy research, and are poised to drastically interfere with naked-eye stargazing worldwide should mitigation efforts be unsuccessful.  
Here we provide predictions for the optical brightnesses and on-sky distributions of several satcons, including Starlink, OneWeb, Kuiper, and StarNet/GW, for a total of 65,000 satellites on their filed or predicted orbits. 
We develop a simple model of satellite reflectivity, which is calibrated using published Starlink observations.
We use this model to estimate the visible magnitudes and on-sky distributions for these satellites as seen from different places on Earth, in different seasons, and different times of night.
For latitudes near 50$^{\circ}$ North and South, satcon satellites make up a few percent of all visible point sources all night long near the summer solstice, as well as near sunrise and sunset on the equinoxes.
Altering the satellites' altitudes only changes the specific impacts of the problem.
Without drastic reduction of the reflectivities, or significantly fewer total satellites in orbit, satcons will greatly change the night sky worldwide. 

\end{abstract}


\section{Introduction} \label{sec:intro}

The placement of tens of thousands of communications satellites into low-Earth orbit (LEO) raises multiple concerns for users of the night sky.
These concerns include the impact of satellites on  astronomical observations, with implications for space exploration and fundamental science, as well as impacts on stargazing for recreation or traditional cultural practices. 
The sheer number of proposed satellites, along with their orbits, will alter the sky far beyond pre-2019 levels.  
In light of these changes, astronomers are undertaking multiple national and international policy initiatives regarding protections for the night sky (e.g., the SATCON1 report\footnote{Full report available here: \url{https://aas.org/sites/default/files/2020-08/SATCON1-Report.pdf}}, the International Astronomical Union Dark and Quiet Skies Report\footnote{Full report available here: \url{https://www.iau.org/static/publications/dqskies-book-29-12-20.pdf}}, and \citeauthor{CSAreport} \citeyear{CSAreport}). 
There have also been a number of efforts to characterize the observed brightness distribution of satellites in orbit, as well as the brightness distribution of proposed satellite megaconstellations (satcons).

\cite{mcdowell2020} explores the impact of 12,000 Starlink satellites based on one of Starlink's earlier proposed orbital distributions. 
This work shows that relatively high concentrations of satellites will be in the sky for mid-latitude locations due to the orbital inclinations of the satellites.
\citet{hainaut_williams2020} modelled a wide range of proposed satcons with 26,000 satellites as seen in different wavelengths. 
They focused on fairly low latitudes, where many ``big-glass'' observatories are located. A general finding is that wide-field surveys will be most strongly affected by satellite light pollution.

\citet{TysonLSST} focuses on the effects of satcons on the upcoming Vera C. Rubin Observatory's Legacy Survey of Space and Time (LSST), finding that if satellites are brighter than $V\simeq7$, they will produce cross-talk within the CCD chips and make observations unreliable across several chips, not just within a streak of pixels.  
It is clear from these previous simulations that satcons have the potential to cause major disruptions to observational astronomy.

Predicting satellite brightnesses prior to launch is difficult. 
A complete model would require extremely detailed information about the shapes and materials in satellites, and launching companies do not provide this information publicly, in general.  
However, as more satellites are launched, observations are being collected to characterize the brightnesses, and this information can be used to characterize generalized reflection models to approximate future satellite brightnesses, or to assess satellite impacts in a statistical sense.
In particular, \citet{horiuchi2020} and \citet{tregloan-reed2021} characterize Starlink's initial voluntary mitigation attempt (``Darksat''), and \citet{Mallama2021,Mallama_lotsa} and \citet{BoleySats} have measured the brightness of original Starlink satellites and their second-generation mitigation attempt of ``Visorsats''.
The SATCON1 report also contains a large number of recent Starlink observations.
All of these observations show that astronomers are rightly concerned about the future of the night sky: the satellites currently in orbit are bright.  Many are naked-eye visible when sunlit, and their brightness is currently unregulated.

In this work, we model brightnesses and the on-sky distribution for a possible future with 65,000 satellites on realistic orbits based on credible filings from satellite operators.  
While we do not know the actual fraction of the proposed satellites that will be launched, the ongoing emergence of new operators, sometimes backed by states, suggests that models exploring tens of thousands of satellites are reasonable. 
We first build a realistic orbital model to find which satellites will be sunlit and potentially visible from different places on Earth, at different times of night, and different seasons (Section~\ref{sec:orbits}).
We then use an assumed satellite reflection model calibrated to recent observations of Starlink satellites  (Section~\ref{sec:satreflect}) to build all-sky visibility predictions (Section~\ref{sec:satvis}), finding that latitudes near 50$^{\circ}$N and S will be the regions most strongly affected  by satellite light pollution.
We briefly explore the effects of satellite altitude (Section~\ref{sec:altitude}), finding that there is no ``best'' satellite altitude; rather, it is a choice between different impacts.
We then discuss in more detail some of the negative effects this will have on observational astronomy (Section~\ref{sec:disc}).
Our satellite modelling code is publicly available so that others may make predictions for their own night sky, and modify it to use with future satellite configurations.

\section{Orbit Model and Sunlit Satellite Distribution} \label{sec:orbits}

We compiled a list of proposed orbits from filings submitted to the U.S.~Federal Communications Commission for Starlink, OneWeb, and Kuiper.  
The information for StarNet/GW is taken from filings at the International Telecommunications Union (ITU). The FCC and ITU filings contain information on the orbital inclination and orbit altitudes, but not necessarily on the details of how satellites are distributed throughout their orbits.  
To build our model, we dispersed the satellites so that they would have a smooth distribution without significant clumping or precisely regular spacing. Two approaches were undertaken in the development of the simulation tools.

\begin{figure}
\includegraphics[scale=0.33]{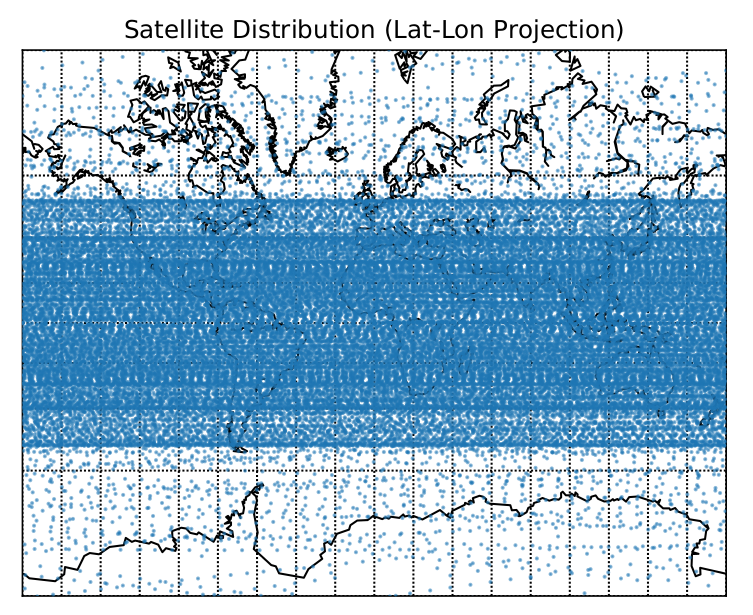}\includegraphics[scale=0.5]{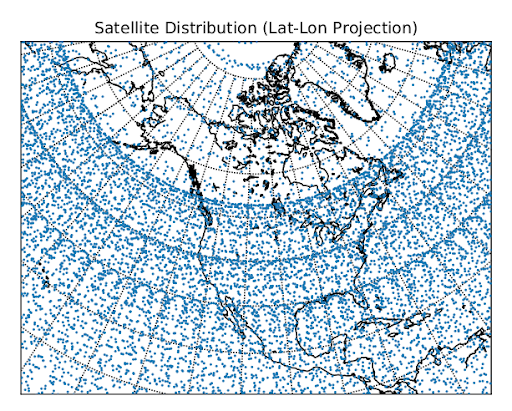}
\caption{A snapshot of the positions of satellites in our model, over the entire globe (left) using a Mercator projection and the slice over North America (right) using a Lambert conformal map projection.  Note that there are obvious bands of higher on-sky density where large numbers of satellites have the same inclination and, as seen in latitude, librate at those parallels. The satellites here use the even-spacing initialization.
\label{fig:ohno}}
\end{figure}

The first, which we call `even-spacing', is as follows: (1) If the filing contains a set number of planes and satellites per inclination plane, then the planes' nodes are evenly distributed from 0 to 360$^{\circ}$ and the satellites in each plane are evenly distributed in mean anomaly, plus a small random offset to avoid excessive clumping due to perfectly regular spacing.
(2) If the filing places all satellites onto their own planes or if the plane distribution is ambiguous, then we evenly space the orbital nodes for each satellite and set the satellites' mean anomalies to ensure a relatively smooth and non-clumping distribution. 
This is achieved by creating an array of evenly spaced mean anomalies between 0 and $2\pi$ for all satellites ($N_{\rm Sat}$) in a given orbital shell and then assigning each satellite the array value that is $\sqrt{N_{\rm Sat}}$ away from the previous satellite's value. This is used for Figure \ref{fig:ohno}.

However, due to the large number of satellites being considered, a second approach is also employed for the webapp (see Section~\ref{sec:satvis}) and related figures. Namely, a quasi-random distribution is used by evenly spacing the orbital nodes of all the planes or, if specific planes are not given, the nodes of all satellites in a shell, and then selecting a satellite mean anomaly from a uniform random distribution on the interval $[0,2\pi)$. 
We refer to the this method as the `quasi-random' initialization. 
Randomizing the satellites in mean anomaly is not an unrealistic approximation.
Starlink in particular has already announced their intentions to use autonomous collision avoidance by its satellites.
In general we expect the distributions to be continuously changing, and fortunately, the overall results are independent of the exact initialization procedure used. Most figures here use this quasi-random distribution.

The altitudes and inclinations of the four different satcons used here are given in Table~\ref{tab:orbits}. 
For simplicity, we list the total number of satellites at a given inclination rather than the breakdown per plane. 
Additional details are in the publicly available code (Section~\ref{sec:satvis}).

\begin{deluxetable}{lccc}								
\tablecaption{Megaconstellation orbital parameters\label{tab:orbits}}								
\tablewidth{0pt}								
\tablehead{								
\colhead{operator} & \colhead{$N_{\rm Sat}$} & \colhead{inclination [$^{\circ}$]} & \colhead{Altitude [km]} 							
}								
\startdata								
Starlink\tablenotemark{a}	&	7178	&	30	&	328	\\	
(USA)	&	7178	&	40	&	334	\\	
	&	7178	&	53	&	345	\\	
	&	2000	&	96.9	&	360	\\	
	&	1998	&	75	&	373	\\	
	&	4000	&	53	&	499	\\	
	&	144	&	148	&	604	\\	
	&	324	&	115.7	&	614	\\	
	&	2547	&	53	&	346	\\	
	&	2478	&	48	&	341	\\	
	&	2493	&	42	&	336	\\	
	&	1600	&	53	&	550	\\	
	&	1584	&	53.2	&	540	\\	
	&	720	&	70	&	570	\\	
	&	348	&	97.6	&	560	\\	
	&	172	&	97.6	&	560	\\	\hline
OneWeb\tablenotemark{b}	&	720	&	87.9	&	1200	\\	
(UK)	&	1764	&	87.9	&	1200	\\	
	&	2304	&	40	&	1200	\\	
	&	2304	&	55	&	1200	\\	\hline
StarNet/GW\tablenotemark{c}	&	480	&	85	&	590	\\	
(China)	&	2000	&	50	&	600	\\	
	&	3600	&	55	&	508	\\	
	&	1728	&	30	&	1145	\\	
	&	1728	&	40	&	1145	\\	
	&	1728	&	50	&	1145	\\	
	&	1728	&	60	&	1145	\\	\hline
Kuiper\tablenotemark{d}	&	1156	&	51.9	&	630	\\	
(USA)	&	1296	&	42	&	610	\\	
	&	784	&	33	&	509	\\	
\enddata	
\tablenotetext{a}{Assumes Starlink Gen1 \citep{FCCStarlinkInitial,FCCStarlinkVLEO,FCCStarlinkGen1Mod} and Gen2 \citep{FCCStarlinkGen2} operate together.}
\tablenotetext{b}{Includes Phases 1 and 2 \citep{OneWebPhase2}.}
\tablenotetext{c}{Based on consolidation of ITU filings \citep{press2021}.}
\tablenotetext{d}{Based on \cite{Kuiper}.}
\tablecomments{Orbits and numbers for Starlink, OneWeb, and Kuiper are listed as filed and amended with the FCC. StarNet/GW is based on ITU filings.}								
\end{deluxetable}

Figure~\ref{fig:ohno} shows a snapshot of the positions of these 65,000 satellites over the world, and over a small region of the globe (centred on North America), to give a more intuitive sense of what the distribution of this extremely large number of satellites looks like. 

Several conspicuous features form, including regions where the satellites have higher densities. 
These are the orbital caustics, caused by the satellites following their inclined orbits and passing through the turnaround points in latitude. 
Multiple caustics are visible in Figure~\ref{fig:ohno}, for example over southern Canada, the middle latitudes of the United States, and northern Mexico.
The locations of these caustics, resulting from a satellite operator's choice in the inclination distribution, has an important consequence for the number of satellites in the sky at different latitudes.

\subsection{Sunlit satellite distribution} 

Now that we have a distribution of satellites in space, we need to determine the number of those satellites that are sunlit for a given observer at a given Solar hour and time of year. 
To proceed, we use an initial coordinate system that places Earth's equator in the X-Y plane and the spin direction aligned in the Z direction. 
The satellites' Cartesian positions are then determined in this coordinate system from their orbital elements. 

The observer's Cartesian coordinates on the Earth's surface are determined next, based on the desired latitude and solar hour of interest. 
We set midnight for an observer to be when they are on the +X axis.

With the position of the observer now set relative to the satellites, we rotate all satellites and the observer about the $Y$ axis by an angle consistent with the Sun's declination for the desired time of year, creating a new coordinate system $\rm X', ~Y=Y',~Z'$.
The Sun is on the $\rm -X'$ axis.
To simplify the calculations, we take a satellite to be sunlit if its $\rm Y'-Z'$ projection is greater than the radius of the Earth or if it is in the domain $\rm X'<0$. In other words, we do not include atmospheric refraction (if it was included, it would cause a slightly larger number of satellites to be partially sunlit on the edge of Earth's shadow). 

\begin{figure}
\includegraphics[scale=0.55]{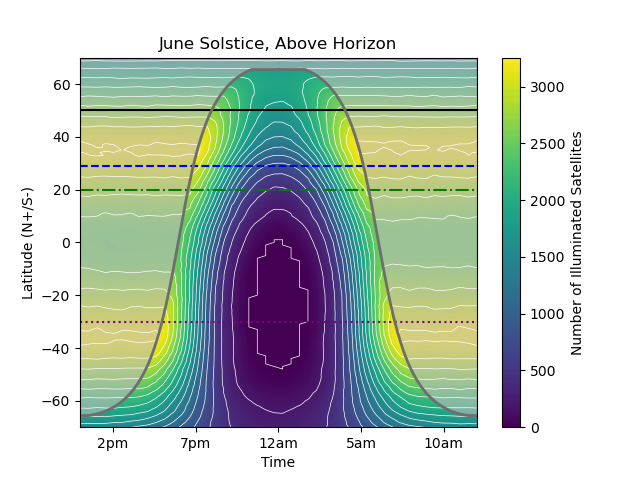}\includegraphics[scale=0.55]{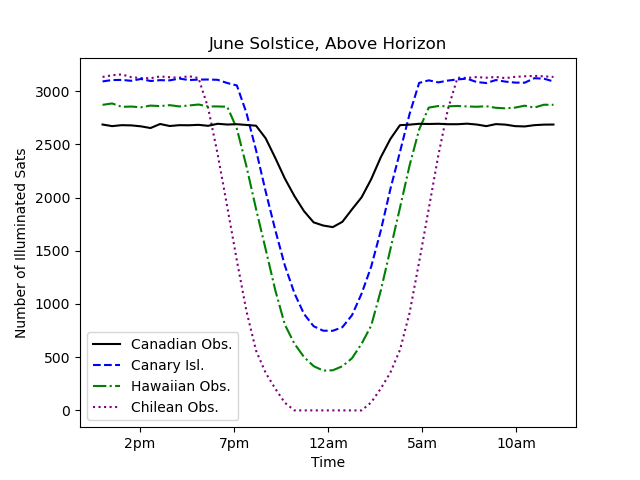}
\includegraphics[scale=0.55]{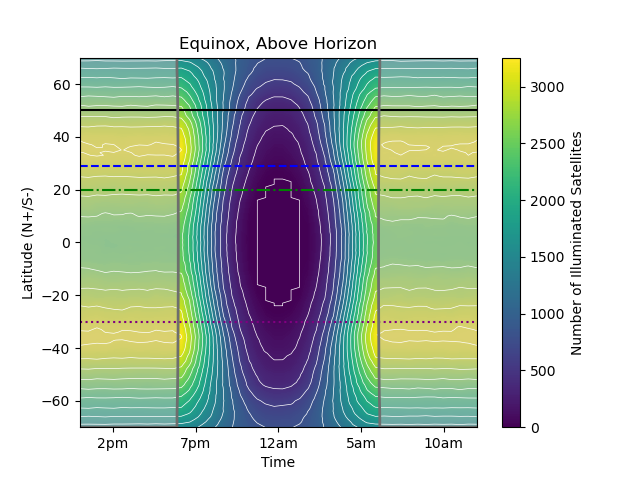}\includegraphics[scale=0.55]{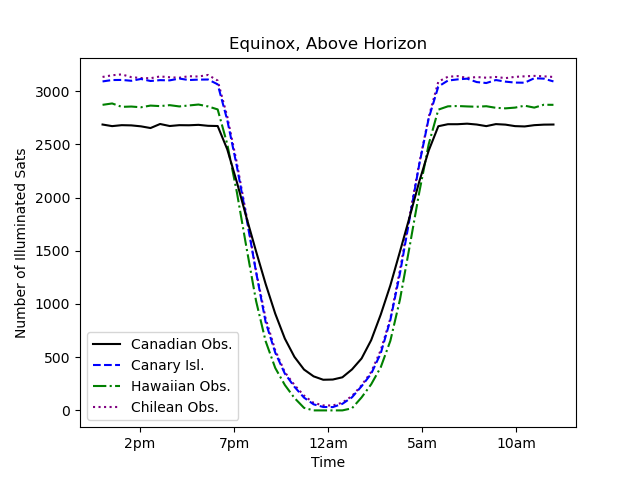}
\includegraphics[scale=0.55]{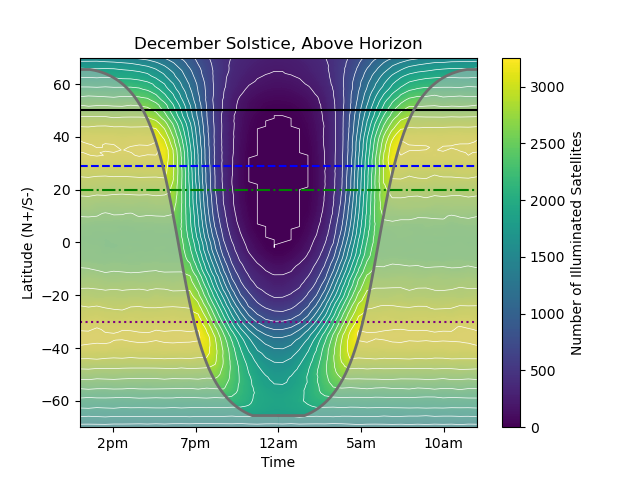}\includegraphics[scale=0.55]{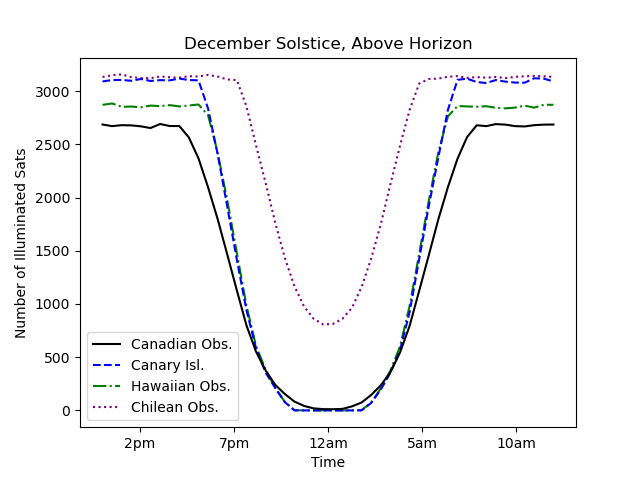}
\caption{The instantaneous number of sunlit satellites above the horizon at different latitudes and different times of night: left panels are heatmaps with daylight (sunrise to sunset) indicated by the grey shaded region; right panels show curves for four different latitudes (shown by lines in the left planels).  The four latitudes highlighted are for the approximate latitude of several observatories in Canada (black solid line), the latitude of observatories on the Canary Islands (blue dashed line), the latitude of observatories on Maunakea in Hawaii (green dash-dotted line), and the approximate latitude of several observatories in Chile (purple dotted line).
\label{fig:sunlit}}
\end{figure}

\begin{figure}
\includegraphics[scale=0.55]{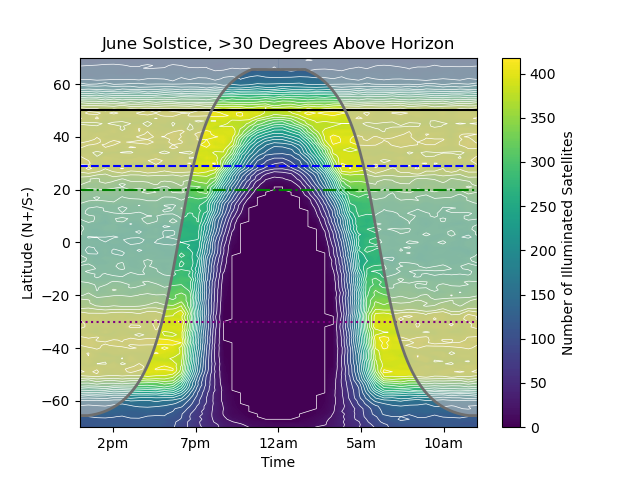}\includegraphics[scale=0.55]{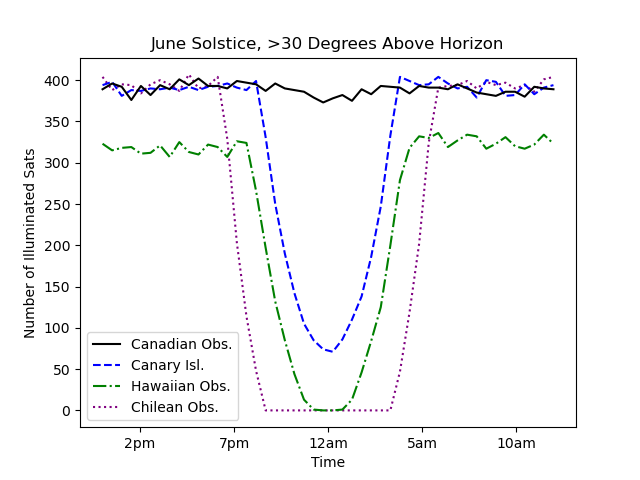}
\includegraphics[scale=0.55]{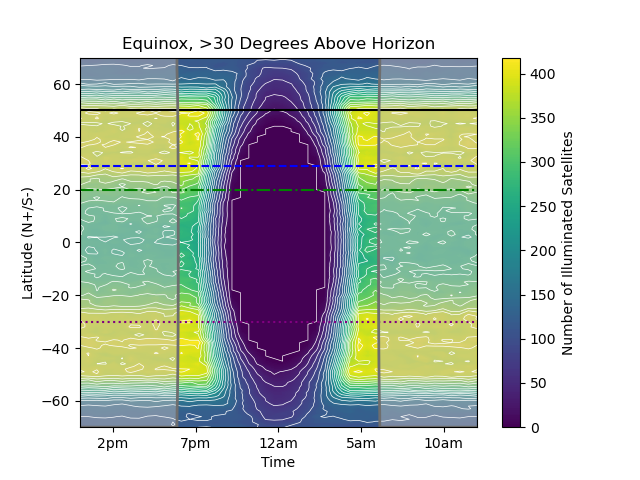}\includegraphics[scale=0.55]{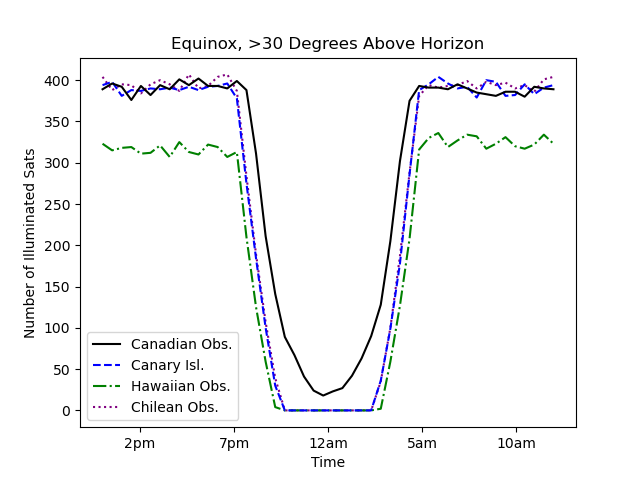}
\includegraphics[scale=0.55]{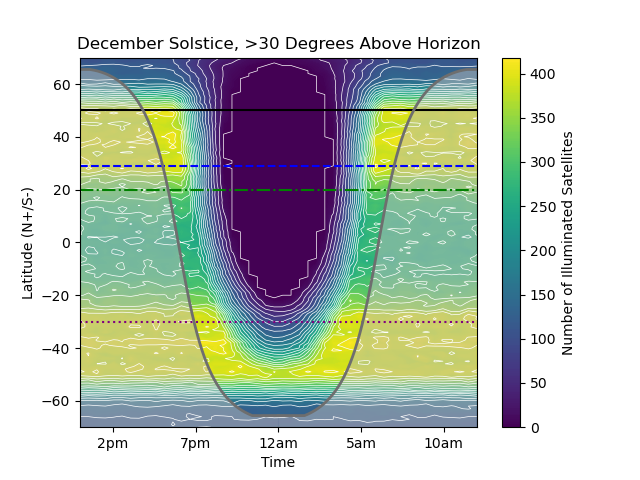}\includegraphics[scale=0.55]{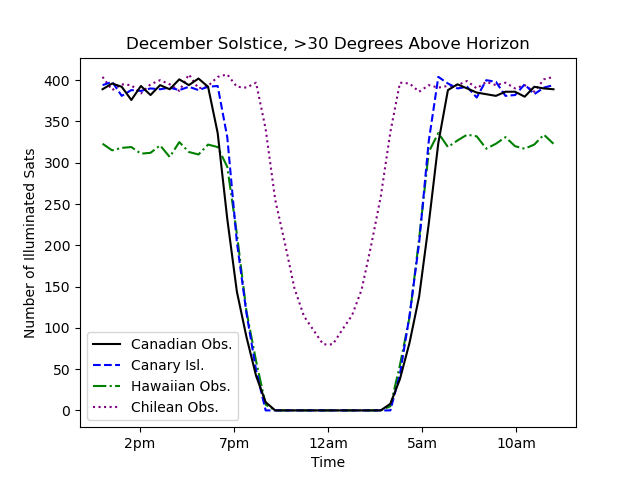}
\caption{The instantaneous number of sunlit satellites more than 30$^{\circ}$ above the horizon at different latitudes and different times of night: left panels are heatmaps; right panels show curves for four different latitudes (shown by lines in the left plots).  Same symbols as Figure~\ref{fig:sunlit}.
\label{fig:sunlit30}}
\end{figure}

Finally, we determine whether a given sunlit satellite is visible to the observer in question, and if so, make cuts based on the satellite's elevation for that observer. We proceed as follows:
Let the vector to the observer be denoted ${\bf r}_{\rm obs}$ and the vector to a given satellite be ${\bf r}_{\rm sat}$, each with an origin at the geocentre. 
The vector from the observer to the satellite is thus
${\bf s} = {\bf r}_{\rm sat}-{\bf r}_{\rm obs}$.
The resulting zenith angle of the satellite for the observer is
$\cos \theta_z = {\bf s}\cdot{\bf r}_{\rm sat}/\left(\big\vert s\big\vert \big\vert {\bf r}_{\rm sat}\big\vert \right)$.

Figure~\ref{fig:sunlit} shows the on-sky density of sunlit satellites at different latitudes, different times of night, and different times of year.
Many of the hundreds or even thousands of sunlit satellites in Figure~\ref{fig:sunlit} are close to the horizon, at higher airmasses than would be used for research observing. 
But there are still hundreds of sunlit satellites more than 30$^{\circ}$ above the horizon (Figure~\ref{fig:sunlit30}), where they could interfere with research astronomy.

In the right panels of Figures~\ref{fig:sunlit} and \ref{fig:sunlit30}, we highlight the number of sunlit satellites over the course of a night from four different observatory latitudes: observatories in Canada, which are all situated close to 50$^{\circ}$~N, observatories on the Canary Islands at 28$^{\circ}$~N, observatories in Hawaii at 20$^{\circ}$~N, and observatories in Chile which are all close to 30$^{\circ}$~S. 

The first feature that is immediately obvious is that latitudes near 40$^{\circ}$~N and S have the most sunlit satellites, consistent with that found by, e.g., \citet{mcdowell2020}.
This is not surprising given the distribution of satellites in Figure~\ref{fig:ohno}.
But due to Earth's shadow, there is a period of few-to-no sunlit satellites for $\sim$5-6 hours around midnight during the winter at these latitudes. 
However, during summer, there are more than a thousand sunlit satellites above the horizon all night long (see 50$^{\circ}$N curves and yellow-filled contours in Figure~\ref{fig:sunlit}, which represents $\sim$3000 sunlit satellites). 
Closer to the equator, where many research observatories are located, the seasonal variation is not so extreme, and there is a period of about 3 hours in the winter and near the equinoxes with few-to-no sunlit satellites.
But there are still hundreds of sunlit satellites all night at these latitudes in each hemisphere's summer. 

The distribution of sunlit satellites more than 30$^{\circ}$ above the horizon is similar (Figure~\ref{fig:sunlit30}).  
Mid-latitudes have a $\sim$5-6 hour period of no satellites centred around midnight in their winter and equinoxes. 
But in the summer, even from these mid-latitudes, there are dozens of sunlit satellites $>$30$^{\circ}$ above the horizon.
This is more extreme at latitudes near 50$^{\circ}$:
there is no drop  in the number of sunlit satellites $>$30$^{\circ}$ above the horizon, even at midnight, at 50$^{\circ}$N or S latitude in their respective summers.

\section{Satellite Visibilities} \label{sec:vis}

Just because a satellite is sunlit does not necessarily mean that it will be visible by eye or even in a telescope. 
However, many of the satellites that are currently being launched did not account for light pollution during the design phases, resulting in satellites that are quite bright at operational altitude.
While satellite reflectivity is now being considered by some operators, brightness reduction efforts thus far have been limited to retrofits of existing designs.
Moreover, there are no regulations concerning how bright a satellite can be -- astronomers are currently dependent on good-faith cooperation with the satellite operators, with the actual brightness and variability of satellites ultimately not known until they are deployed and observed.

\subsection{Satellite Reflection Models} \label{sec:satreflect}

A satellite's brightness depends on its non-trivial bidirectional reflectance distribution function (BRDF), which takes into account the detailed shape  of the satellite and its materials. The brightness at any given moment may depend sensitively on the satellite's exact orientation in space with respect to the incoming light and the observer.  
Satellites are known to vary in brightness as they move across the sky, continuously changing their orientation with respect to the observer.
This evolving perspective can lead to sudden, and in the case of flares, dramatic changes in brightness.
This brightness variability can be observed, in principle, even for a single satellite traversing the field of view of a telescope. 

Satellite variability and flares have long been observed by astronomers, both by purposefully observing satellites and by the accidental passage of satellites through research images \citep[e.g.,][]{hainaut_williams2020,Mallama2021}; naked-eye observations by many stargazers around the world also note this behaviour.\footnote{This includes the authors of this paper.}

In practice, the detailed BRDF is typically unknown and simplifying assumptions must be employed. Such assumptions will not capture sudden satellite variability, but can still be useful enough to simulate the overall trends and distributions of satellite brightness.

We use a diffuse Lambertian sphere reflection model \citep[e.g.,][]{pradhan_2019} to estimate the distribution of visible magnitudes of these sunlit satellites:
\begin{equation}
m_{v} = m_{\rm Sun} - 2.5\log \left(\frac{2}{3\pi^2} A \rho [ (\pi - \phi)\cos\phi + \sin\phi ] \right) + 5\log R
\end{equation}
where $m_{\rm Sun}$ is the apparent magnitude of the Sun \citep[-26.77 in $V$-band, -26.47 in $g'$-band for AB magnitudes;][]{willmer2018}, $A$ is the cross-sectional area of the satellite, $\rho$ is the Bond albedo, $\phi$ is the Solar phase angle between the observer and the satellite, and $R$ is the distance between the observer and the satellite.
We do not know the reflecting area  or albedo of future satellites, and cross-sectional area will vary depending on viewing phase and satellite attitude. 
To further simplify our reflectance model, $\rho$ and $A$ are combined into one albedo-area variable, the effective area $\zeta = \rho A$.

To calibrate our model, we compared our model satellite brightness distribution with the observed distribution of Starlink satellites magnitudes from \citet{BoleySats}, which used the Plaskett Telescope at the Dominion Astrophysical Observatory in Victoria, BC (latitude 48.5~$^{\circ}$N) to observe a set of Starlink satellites.
While OneWeb satellites have so far proved fainter than Starlinks, they have approximately the same absolute magnitude \citep{Mallama2021}.  Since Kuiper and Starnet/GW do not yet have any satellites in orbit, a set of Starlink observations is a reasonable starting point for our model calibration.
We set up the model using the same general observing directional biases, matching the season, time of night of the observing run, and the latitude, and we only used Starlink satellite orbits near 550~km altitude, since that is the most prominent orbital shell occupied as of the time of observations (see Table~\ref{tab:orbits}). Such Starlink satellites have orbital inclinations of about $53^\circ$, meaning their orbits can pass through the sky at low zenith angles at the latitude of the Plaskett Telescope. 

As discussed in \citet{BoleySats}, we first attempted to use only the Visorsat Starlinks in our calibration, in hopes that these brightness mitigations would be used on all future satellites. 
But upon further investigation, we found that although all Starlinks since the 7 August 2020 launch are reportedly equipped with visors\footnote{see, e.g., \url{https://directory.eoportal.org/web/eoportal/satellite-missions/s/starlink}}, there is no available information on how many of those satellites actually have the visors successfully deployed.
The variation in brightness observed in \citet{BoleySats} leads us to speculate that there may be inconsistencies among the visors, or that their effectiveness is very sensitive to the precise orientation between the satellite, Sun, and observer. We note that the latter case is hinted at by observations in \citet{Mallama_lotsa}.

\begin{figure}
\begin{centering}
\includegraphics[scale=0.55]{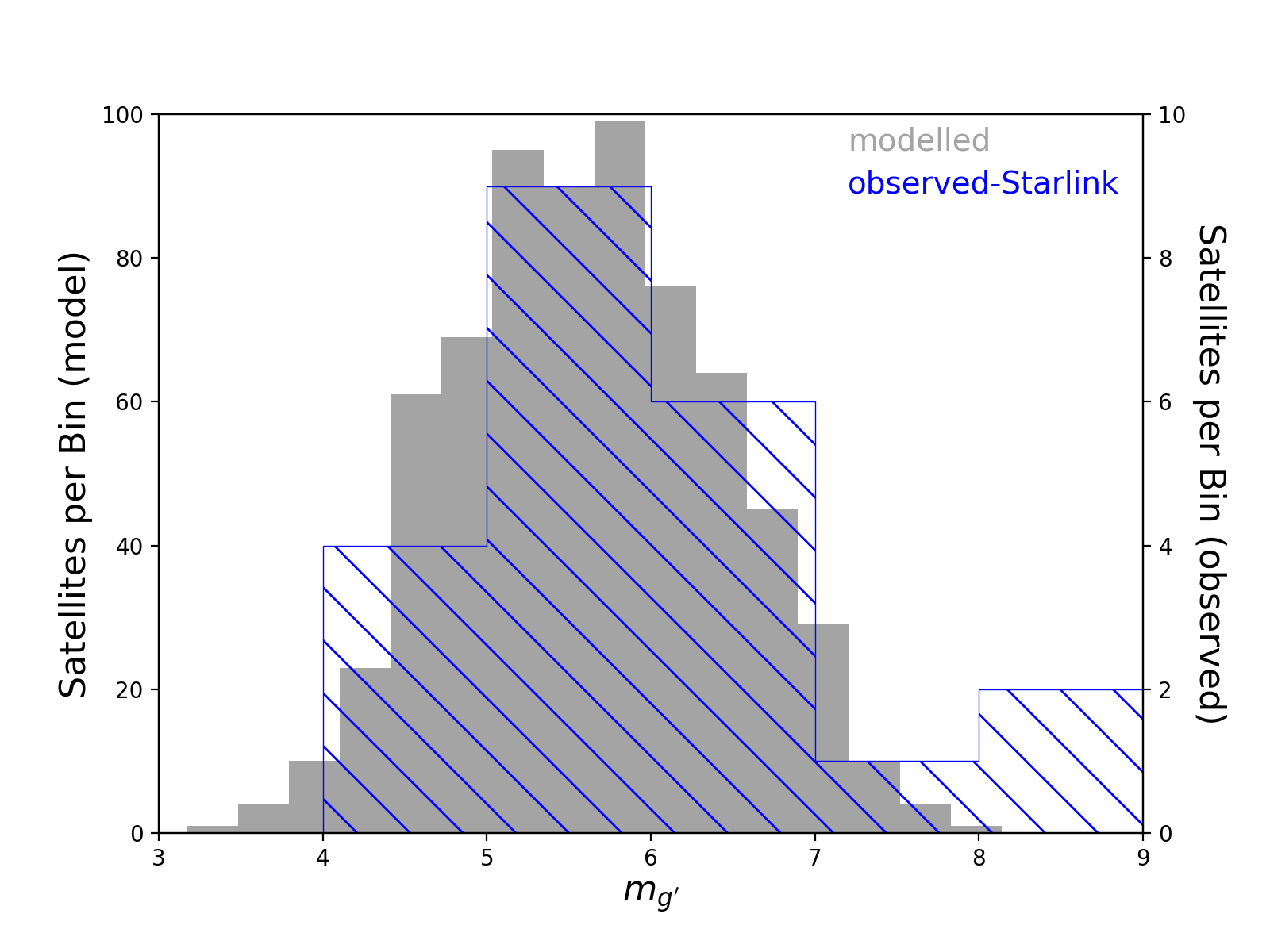}
\caption{Histograms of $g'$ magnitudes for observations of Starlink satellites (blue hatched) compared with our satellite model output (grey). 
The observations \citep{BoleySats} were obtained in July 2021 from the Plaskett 1.8 m telescope at the Dominion Astrophysical Observatory (latitude 48.5$^{\circ}$N). 
The satellites were observed between $\pm 1.5$ hr of midnight. The Sun's declination at the time of the observations was about 21$^\circ$. Only Starlink satellites near the 550 km orbital shell were targeted.
This figure demonstrates that using an effective surface area of $\zeta$~=~0.8~m$^2$ is a reasonable choice for our model, with the important caveat that this is applicable to a mixture of Starlink satellites: future satellites may have a very different actual brightness distribution depending on the size, shape, and materials used.
\label{fig:sathist}}
\end{centering}
\end{figure}

With the above in mind, we fit our model to the entire distribution of observed Starlink satellites, both listed Visorsats and original-design Starlinks, which is shown in Figure~\ref{fig:sathist}.  
This may be a more realistic distribution of satellite brightnesses, since it includes both satellites with mitigation (Visorsats) and those without (original Starlinks). 
]We also note that initial observations of Visorsats and original Starlinks by amateur astronomers \citep[e.g.,][]{mcdowell2020} showed a wide range of brightnesses for both satellite types.
Without regulatory obligations to make their satellites fainter, or widely adopted and  followed industry standards, there will likely be a large range of reflectivities in future satellites.

The modelled albedo-area variable was changed until the resulting model distribution of Starlink satellites matched the observed distribution from \citet{BoleySats}, under similar observing directional and time biases (Figure~\ref{fig:sathist}).  
We also added a small scatter in brightness (Gaussian noise with $\sigma=0.5$ magnitudes) to simulate the scatter observed in \citet{BoleySats}.
This best fit corresponds to a satellite effective area $\zeta\approx0.8~\rm m^2$, which could realistically be matched by a reflecting satellite area of 4~m$^2$ with an albedo of 0.2, as an example.

The Plaskett measurements compare favourably with measurements from \citet{Mallama2021}, but taken from 48.5$^{\circ}$N rather than 39$^{\circ}$-44$^{\circ}$N, where satellites will remain bright throughout the night for the time of year of the observations (Section~\ref{sec:satvis}). The median observed $g'$-magnitude in the \citet{BoleySats} sample, which includes Starlink satellites with and without visors, is $g'=5.7$, compared with a median of 5.9 measured by \citet{Mallama2021}, which used an observing filter within 0.1 mags of $V$-band.  
For the Sun, $g'-V = 0.3$, so $V\approx 5.9$ corresponds to $g'\approx 6.2$.
However, the observational biases between the samples are expected to be different. 
The results presented by a follow-up study \citep{Mallama_lotsa} find a median $V$ of about 6.8 for Visorsats, after they have been adjusted to a range of 1000~km. Correcting this to 550~km, yields a 550~km ``absolute’’ $V$-magnitude of $H_V\approx 5.5$ or $H_{g’}\approx 5.8$. This compares favourably to the Visorsat-only subsample from \citet{BoleySats} which measured $H_{g’}=5.7$ (coincidentally the same value as the apparent magnitudes of their full sample).

We reiterate that there are many reflection effects that we are ignoring in our satellite brightness model, and that the brightness distribution is sensitively dependent on effective cross-sectional area and exact observing geometry.
But based on comparison between this model and published observations of one satcon (Starlink), we find that the overall scale of the satellite brightness is well-represented.

\subsection{Visible magnitude predictions} \label{sec:satvis}

We now apply this model calibration to the satellite distributions to calculate their apparent magnitudes as seen in the sky at different latitudes, different times of year, and different times of night, with the caveats discussed above.
We also include realistic atmospheric extinction in this model \citep{extinct1989}, which is significant for the lowest elevation satellites. 
We calculate the visual magnitudes for all sunlit satellites in each field of view, but we pay particular attention to three magnitude cuts: those with $g<5$, which are visible from a typical suburban environment; those with $g<6.5$, which are naked-eye visible from a truly dark sky; and those with $g<7$, which are bright enough to cause serious data loss for LSST \citep{TysonLSST} and other research facilities.  

The source code needed to reproduce all of these brightness simulation figures is available at \url{https://github.com/hannorein/megaconstellations/}.
Furthermore, we provide an easy-to use web interface to make predictions at custom observatory locations, times of night, and different seasons at \url{http://megaconstellations.hanno-rein.de}. 

\begin{figure}
\resizebox{0.99\columnwidth}{!}{\includegraphics{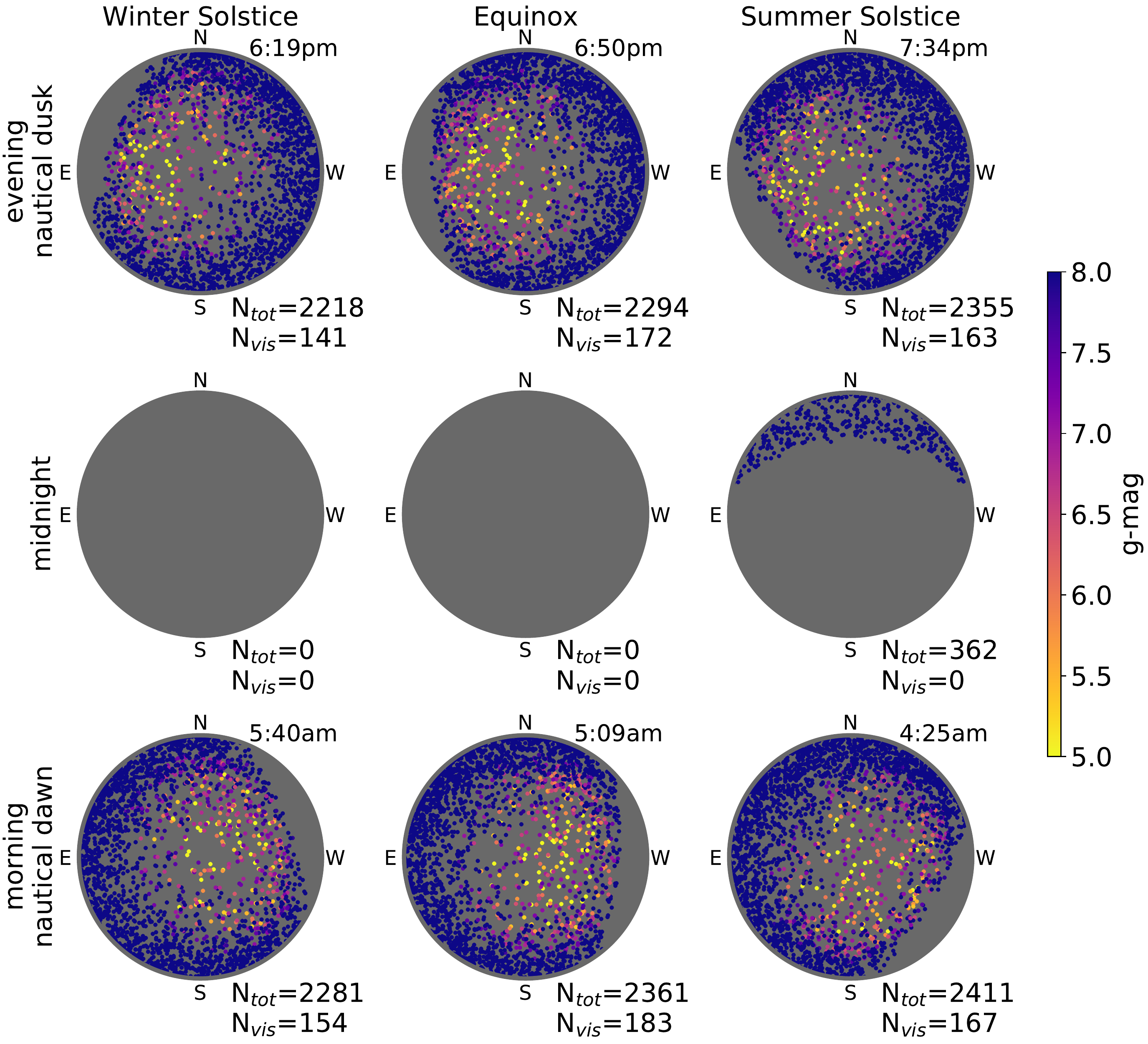}}
\caption{All-sky snapshots of the predicted satellite distribution and brightnesses as viewed from Hawaii (latitude 20$^{\circ}$~N, will be the same distribution at 20$^{\circ}$S, but flipped N-S on the sky).  Colour shows $g$-band magnitude.  Moving left to right across the grid: Winter solstice, equinox, summer solstice.  Top to bottom across the grid: Nautical dusk, midnight, nautical dawn. For each snapshot, N$_{tot}$ is the total number of sunlit satellites in the sky from this location at this time of day and year, and N$_{vis}$ is the number of satellites that are naked-eye visible (approximately $g<6.5$) from this location at this time of day and year.  Note that these numbers are not exact and are expected to vary by $\sqrt{N}$ due to randomization of satellite distribution within their orbital shells.  The satellite distribution is faint and close to the horizon or not illuminated at all for most of the night, except for near dusk and dawn, when the brightest satellites will be close to the zenith, with hundreds being naked-eye visible.
\label{fig:hawaiisats}}
\end{figure}

\begin{figure}
\resizebox{0.99\columnwidth}{!}{\includegraphics{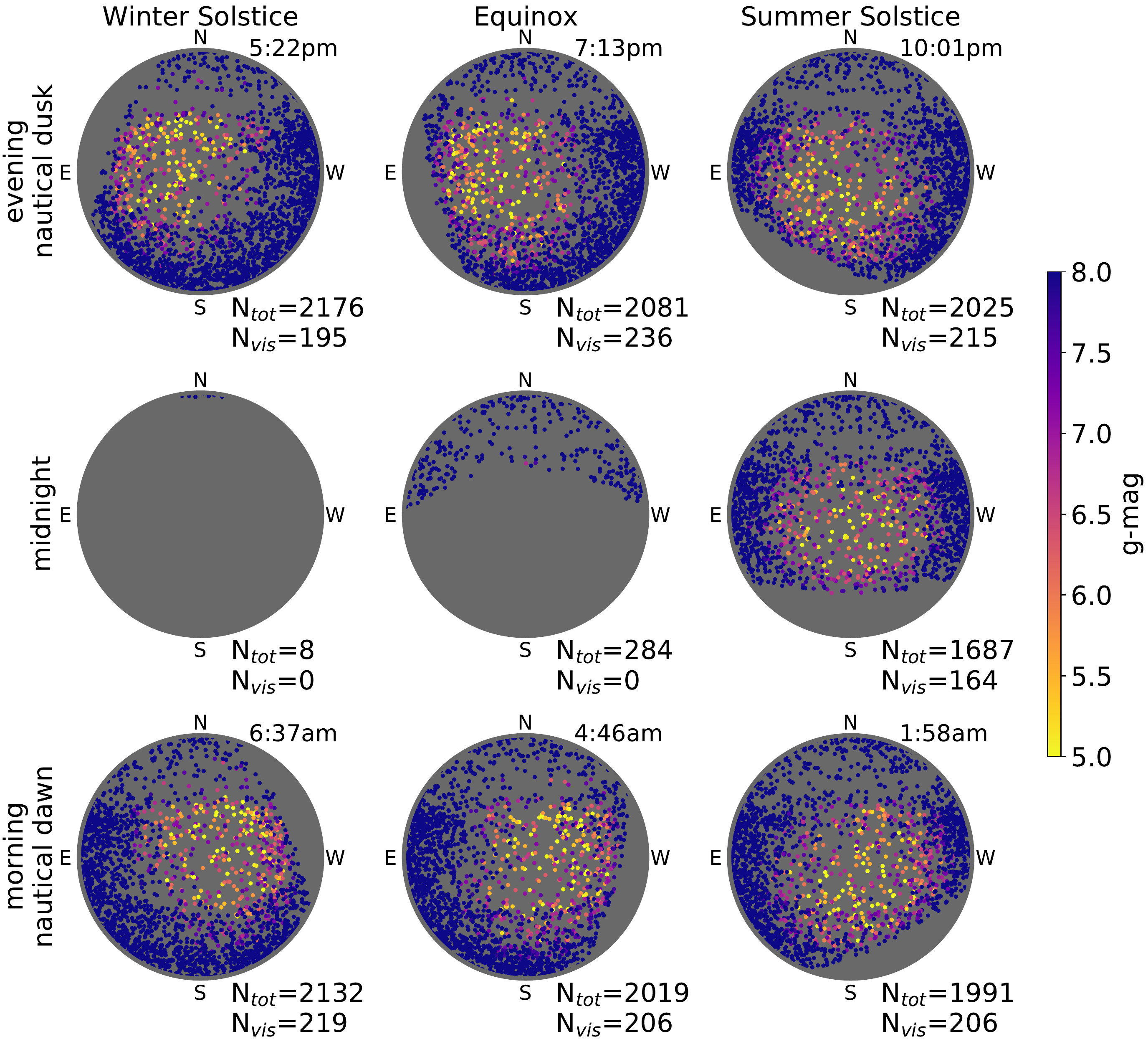}}
\caption{Same plots as Figure~\ref{fig:hawaiisats}, but viewed from a latitude of 50$^{\circ}$~N (e.g., many Canadian observatories).  In the summertime, the number of sunlit and visible satellites (hundreds) only drops by a small fraction over the course of the night.  
\label{fig:canadasats}}
\end{figure}

\begin{figure}
\resizebox{0.99\columnwidth}{!}{\includegraphics{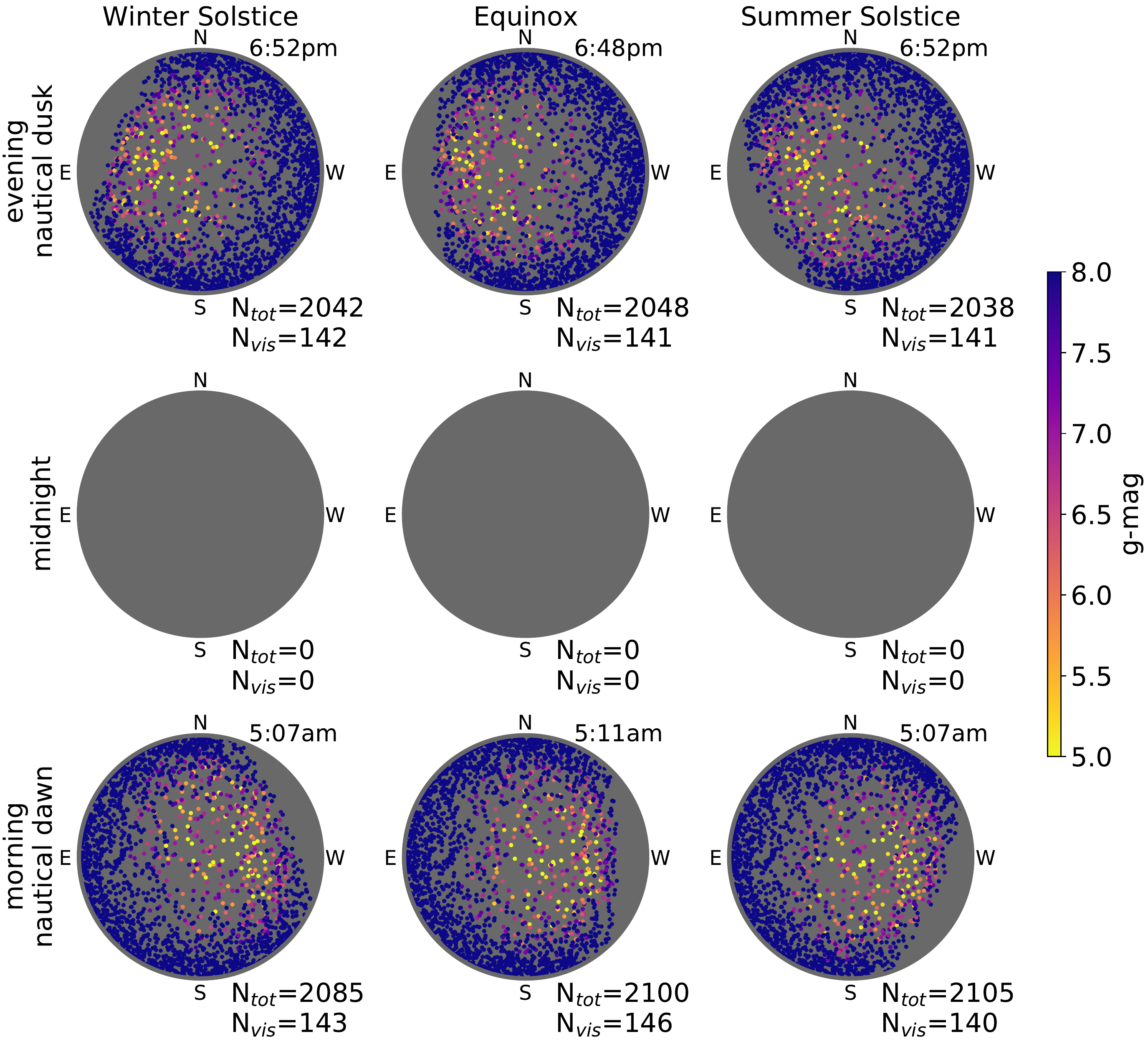}}
\caption{Same plots as Figure~\ref{fig:hawaiisats}, but viewed from the Equator (latitude 0$^{\circ}$).  There are no sunlit satellites at midnight in any season.
\label{fig:eqsats}}
\end{figure}

\begin{figure}
\resizebox{0.99\columnwidth}{!}{\includegraphics{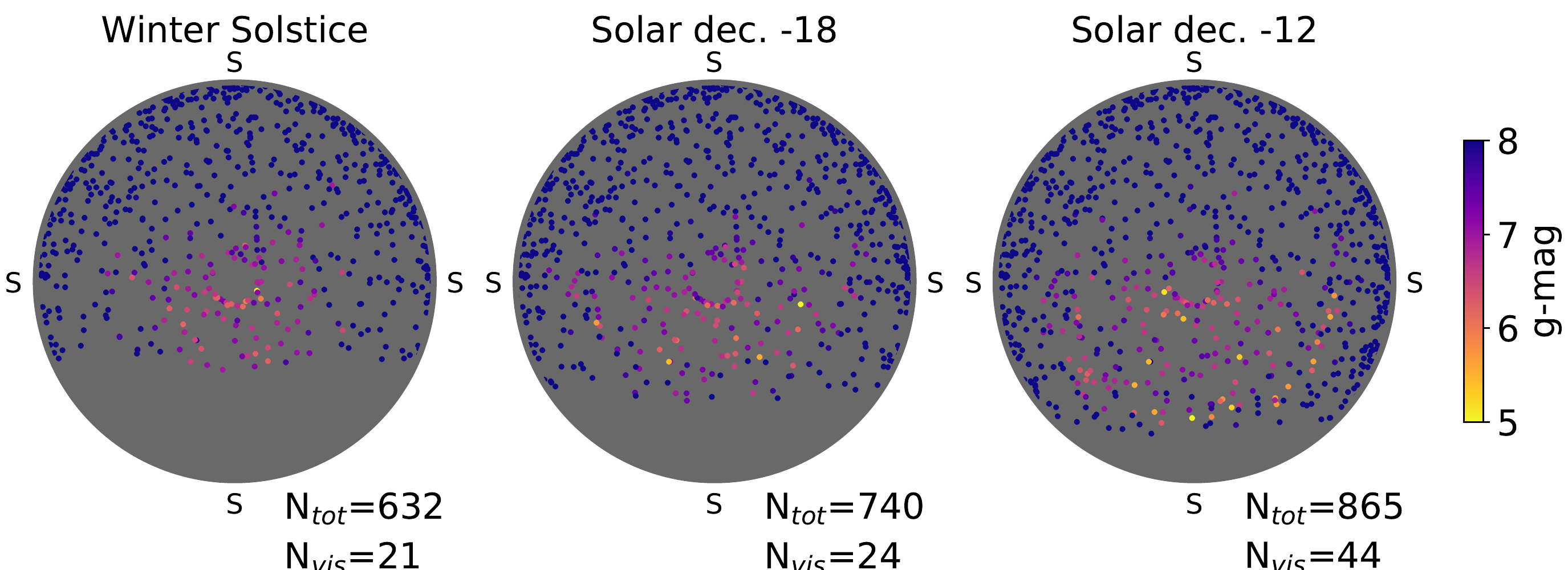}}
\caption{Similar to Figure~\ref{fig:hawaiisats}, but viewed from the North Pole (latitude 90$^{\circ}$N).  Because of extreme seasonal illumination, satellites would not actually be visible close to the summer solstice due to 24-hour daylight, or close to the equinox because of 24-hour near-daylight, but the dozens of satellites brighter than $g<6.5$ will be visible 24 hours per day during the winter.  For comparison, we show midnight on the Winter Solstice (24 hours of full darkness), at a Solar declination of -18$^{\circ}$ (24 hours of astronomical twilight), and at a Solar declination of -12$^{\circ}$ (24 hours of nautical twilight).  Note that from the North Pole, all directions are south.
\label{fig:northsats}}
\end{figure}

Figures~\ref{fig:hawaiisats}-\ref{fig:northsats} show the calculated all-sky satellite distributions and brightnesses for a selection of seasons and times of night for different latitudes on Earth.
These are also available in the form of time-lapse movies at \url{http://uregina.ca/~slb861/satcon.html}.  The results are further summarized in Table~\ref{tab:mags}, where the number of satellites above each magnitude cut is compared to the number of visible stars in the observer's sky.  At latitudes near 50$^{\circ}$N and S, where the light pollution is the worst, up to 8\% of all point sources in the sky are satellites at certain times of night.
Because the satellite orbits are symmetric about Earth's equator, the simulations can be applied to either hemisphere for the corresponding season.
For example, the all-sky view for Hawaii (20$^{\circ}$~N) on the June Solstice is equally applicable to S\~{a}o Paolo, Brazil (20$^{\circ}$~S) on the December Solstice, but the all-sky image must be flipped north-south. 
The number of sunlit and visible satellites is given for each sky-field.  These numbers are slightly randomized due to the pseudo-random satellite distribution and the small random scatter added in to the satellite brightnesses. 

\begin{deluxetable}{lllrrrrrr}																				
\tablecaption{Satellite magnitudes \label{tab:mags}}																				
\tablewidth{0pt}																				
\tablehead{																				
\colhead{Latitude}	&	\colhead{Season}	&	\colhead{time}	&	\colhead{$g<5$} 			&	&	\colhead{$g<6.5$} 	&			&	\colhead{$g<7$} 			\\	
	&		&		&	\colhead{number}	&	\colhead{\% sats}		&	\colhead{number}	&	\colhead{\% sats}		&	\colhead{number}	&	\colhead{\% sats}	}
\startdata																				
60$^{\circ}$N/S & summer  & midnight           &  19 & 2\% & 292 & 6\% & 480 & 6\% \\ 
                & equinox & midnight           &   0 & 0\% &   9 & 0\% &  16 & 0\% \\ 
                &         & nautical dusk/dawn &   9 & 1\% & 137 & 3\% & 244 & 3\% \\ \hline
50$^{\circ}$N/S & summer  & midnight           &  48 & 6\% & 259 & 6\% & 386 & 5\% \\ 
                &         & nautical dusk/dawn &  60 & 7\% & 317 & 7\% & 487 & 6\% \\ 
                & equinox & midnight           &   0 & 0\% &   0 & 0\% &   0 & 0\% \\ 
                &         & nautical dusk/dawn &  72 & 8\% & 311 & 7\% & 438 & 5\% \\ \hline
40$^{\circ}$N/S & summer  & midnight           &   0 & 0\% &  34 & 1\% &  71 & 1\% \\ 
                &         & nautical dusk/dawn &  79 & 9\% & 327 & 7\% & 487 & 6\% \\ 
                & equinox & midnight           &   0 & 0\% &   0 & 0\% &   0 & 0\% \\ 
                &         & nautical dusk/dawn &  76 & 9\% & 322 & 7\% & 493 & 6\% \\ \hline
30$^{\circ}$N/S & summer  & midnight           &   0 & 0\% &   1 & 0\% &   4 & 0\% \\ 
                &         & nautical dusk/dawn &  78 & 9\% & 299 & 6\% & 450 & 6\% \\ 
                & equinox & midnight           &   0 & 0\% &   0 & 0\% &   0 & 0\% \\ 
                &         & nautical dusk/dawn &  78 & 9\% & 310 & 7\% & 480 & 6\% \\ \hline
20$^{\circ}$N/S & summer  & midnight           &   0 & 0\% &   0 & 0\% &   0 & 0\% \\ 
                &         & nautical dusk/dawn &  60 & 7\% & 246 & 5\% & 375 & 5\% \\ 
                & equinox & midnight           &   0 & 0\% &   0 & 0\% &   0 & 0\% \\ 
                &         & nautical dusk/dawn &  56 & 7\% & 259 & 6\% & 403 & 5\% \\ \hline
\multicolumn{3}{l}{\textbf{Background stars}}& \textbf{804} & & \textbf{4395} & & \textbf{7702} & 
\enddata																				
\tablecomments{``\% sats'' means the percentage of point sources in the sky at that are satellites, not stars, at each magnitude cut, for top-of-the-atmosphere magnitudes (i.e., no extinction).}																				
\end{deluxetable}																																								
There are notable patterns in the on-sky satellite brightness distributions.
As expected, satellites get fainter close to the horizon, and are brightest as they pass overhead - this is due to a combination of effects: the phase angle, satellite range (distance between the satellite and the observer), and the satellite's instantaneous airmass. The satellite will be closest to the observer when it passes through the zenith. 
As the satellite moves toward the Sun from the zenith, its range increases and its phase angle also increases, causing the satellite to become faint quickly under the diffuse sphere model, 
although the results of \citet{Mallama_lotsa} show that this behaviour can be complex. 
Moreover, as the satellite moves away from the zenith and away from the Sun, the range is increasing (causing dimming) while the phase angle is decreasing (causing brightening), two competing effects that make overall on-sky brightness patterns difficult to predict.

The shadow of the Earth is apparent in many of the all-sky images, and is what blocks the closest overhead satellites from being sunlit close to the equator. 
In latitudes closer to the poles, because of geometry, the Earth's shadow does not block the satellites near the zenith, resulting in bright satellites being visible throughout the night.

In the locations close to the equator ($0^{\circ}$, Figure~\ref{fig:eqsats}, and Hawaii at 20$^{\circ}$~N, Figure~\ref{fig:hawaiisats}), only faint satellites are visible, primarily close to the horizon.  
From Hawaii, and to a greater extent from the Vera Rubin Telescope site in Chile (30$^{\circ}$~S), there are a handful of bright satellites crossing the zenith close to sunrise and sunset in the summer. 
This could restrict the portions of the sky that are available for scientific studies in the summertime, when the weather is generally better.
This effect of satcons on research astronomy can also be seen in Table~\ref{tab:mags}, where satellites make up only a tiny fraction of all points sources in the sky at midnight in the range of $g$-magnitudes tested at 20-30$^{\circ}$ latitudes. 

At higher latitudes of 50$^{\circ}$~N (Figure~\ref{fig:canadasats}), where many Canadian and European observing facilities are located, as well as a large fraction of the population of Canada, Europe, and northern-tier US states, the problem of light pollution from satellites is much more severe seasonally (and is equally severe at 50$^{\circ}$~S).
In the winter, only a handful of satellites are visible late into the night and are faint, while near the equinox, satellites are visible at midnight, but they are all faint.
However, near the summer solstice, there are many bright satellites crossing the zenith all night long. 
The shadow of the Earth blocks many of the would-be brightly-lit satellites close to the equinox, but this does not occur closer to the summer solstice.
Table~\ref{tab:mags} quantifies this: at 50$^{\circ}$N or S latitude, near the summer solstice, satellites will make up 6-7\% of all visible point sources in the night sky at $g<5$, $g<6.5$, and $g<7$ within a couple hours of sunrise or sunset;
Restated: as seen by the naked eye, about one in every 14 stars in the sky will actually be a satellite.
Close to midnight, this fraction drops slightly to 5-6\%, but as shown in Figure~\ref{fig:canadasats}, the brightest satellites ($g$-magnitudes near 5-5.5) remain close to the zenith at all hours of the night.

It is nearly as severe of an effect at 40$^{\circ}$N or S: 6-9\% of point sources in the sky are going to be satellites for these magnitude cuts within a couple hours of sunrise and sunset near the summer solstice.
At 40$^{\circ}$N or S, the brightness of satellites drops off more steeply toward midnight and toward the equinox as compared with 50$^{\circ}$N or S, so many fewer satellites are visible at midnight.

At even more northerly/southerly latitudes of 60$^{\circ}$N or S, the fraction of satellites is 2-6\% all night long near the summer solstice. The fraction drops quickly close to the equinox.  Of course, close to the summer solstice at these high latitudes, there is very little darkness, so the summer light pollution by satellites might be much less noticeable than at latitudes of 50$^{\circ}$ or even 40$^{\circ}$.

As a check of our calculations, and the most extreme case, we also include the view from the North Pole (Figure~\ref{fig:northsats}, equally applicable to the South Pole).
Interestingly, there are never any satellites directly overhead: this is because satellites are not inclined by exactly 90$^{\circ}$.
Of course, the visibility pattern will be quite different at the North Pole than at other latitudes: the Summer Solstice satellites will not be visible at all due to 24-hour daylight, while the Winter Solstice satellites will be visible all day and night due to 24-hour darkness.
Close to the Winter Solstice, only the highest altitude satellites close to the zenith will be illuminated and visible (presumably 24-hours per day).
For comparison, we also modelled the sky view at midnight, with the Sun at a declination of -18$^{\circ}$ (October 24), when there is 24 hours of astronomical twilight, and at -12$^{\circ}$ (November 12), when there is 24 hours of nautical twilight. 
As the Sun creeps to higher declinations, the bright satellites appear further and further south in the polar sky.
While there aren't any observing facilities at the North Pole, there are high-latitude facilities, and these predictions are equally applicable to Antarctic astronomical facilities.

These calculations highlight the unequally distributed light pollution that will result from satcons at different places on Earth: according to our simulations, 50$^{\circ}$N and S latitudes will experience some of the worst effects of light pollution compared to anywhere else on Earth. 

\subsection{The Effects of Altitude} \label{sec:altitude}
\begin{figure}
\centering\resizebox{0.8\columnwidth}{!}{\includegraphics{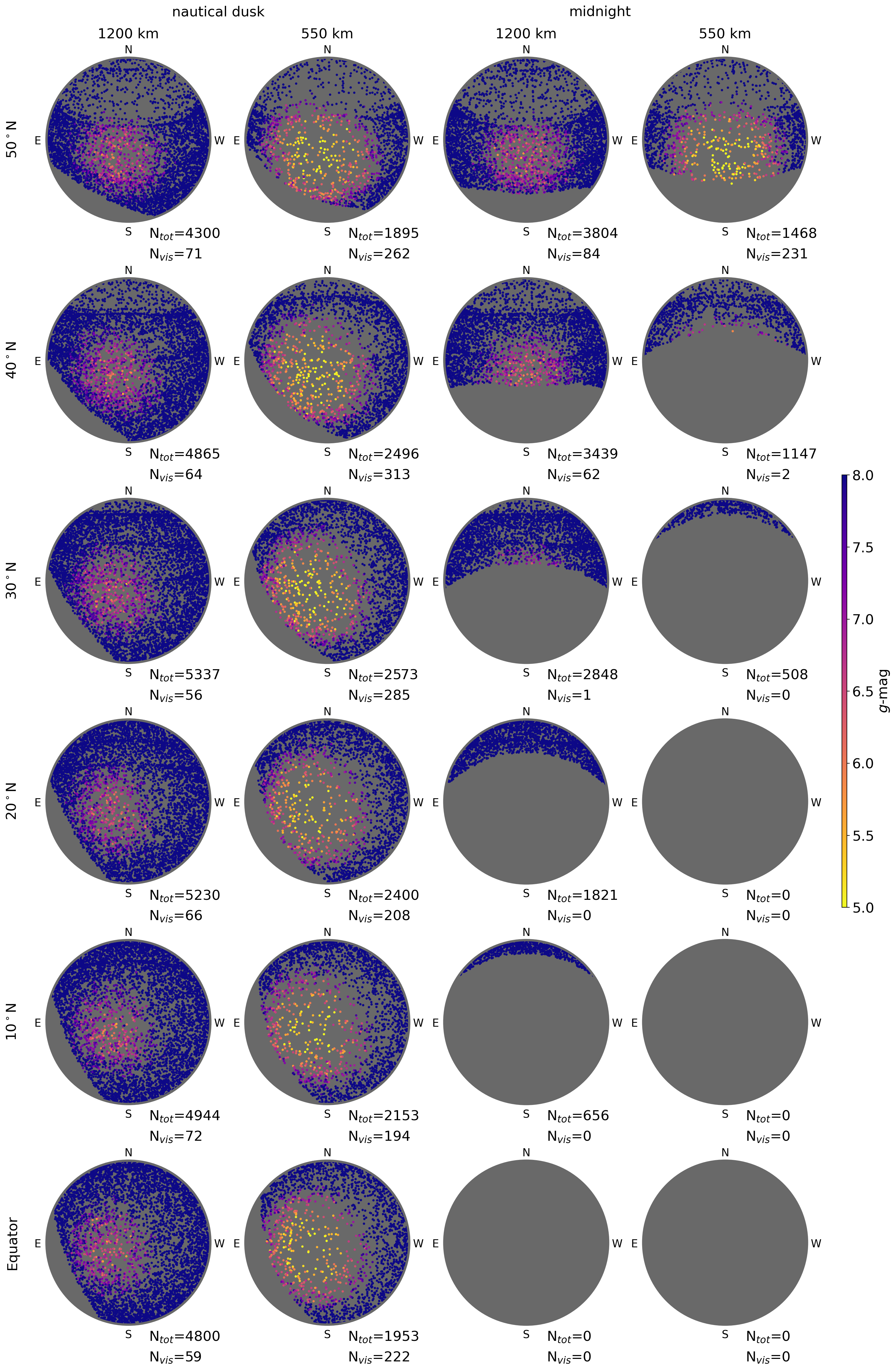}}
\caption{
65,000 satellites on the same orbits as Table~\ref{tab:orbits}, but all at an altitude of 1200~km (first and third columns) or 550~km (second and fourth columns), as viewed on the summer solstice.  
Models are shown at nautical dusk (first and second columns) and at midnight (third and fourth columns).  
Rows show the models at different latitudes.
This plot highlights the fact that there is no ``best'' altitude for satellites, it is a trade-off of different problems for different locations on Earth.
\label{fig:altcompare}}
\end{figure}

One recommendation from the SATCON1 report is that satellites should orbit at lower altitudes. While this makes them brighter, it also causes them to move through a given telescope field of view more quickly, smearing the reflected sunlight across more pixels in a shorter amount of time, decreasing the damage to science images.
Our models here show that while this strategy works well for lower latitudes, where many research telescopes are located, it will lead to many bright satellites visible all night long at higher latitudes during times close to the summer solstice.  

To more thoroughly explore the effects of satellite altitude, we build two toy models.
Each toy model uses the same number of satellites and the orbits from Table~\ref{tab:orbits}, but all of the satellites have been moved to an orbital altitude of 550~km in one model and 1200~km in the other. 
Figure~\ref{fig:altcompare} shows these two satellite altitude toy models on the summer solstice, at nautical dusk and at midnight, and at many different latitudes, illustrating the patterns in light pollution across the Earth caused by satellites at different altitudes.

The number of sunlit satellites in the sky is much larger for the 1200~km model than the 550~km model, at all latitudes.
The satellites in the 1200 km model are also  on average much fainter, with significantly fewer satellites that are visible to the naked eye for all latitudes and at all times of night.
However, while there are fewer bright satellites in the higher altitude model, the rate of motion across the sky is slower, making each satellite visible in a given telescope's field of view for a longer period of time, as noted above. 
With this in mind, satcon operators will require more satellites at lower altitudes than at higher altitudes to provide comparable beam coverage, all other things being equal. 
This is not captured in this toy model, which uses the same number of satellites for the two altitude cases. 
Thus, the total number of satellites in the sky might not vary too much according to the chosen altitude, depending on the satcon designs. 
These trade-offs need to be handled with a high degree of care and continuously re-evaluated. 

Another issue to consider is that while satellites are overall brighter in lower orbits, the number of sunlit and visible satellites drops very quickly after nautical dusk for the lower latitudes, giving a period of very few sunlit satellites for most of the night.
However, this fast drop-off in low-orbit visible satellites does not happen at higher latitudes. For example, at 50$^{\circ}$~N, there continues to be hundreds of naked-eye visible satellites all night long near the summer solstice.
The higher orbit satellite model also produces many naked-eye visible satellites at high latitudes all night long in the summer, but not as many as the lower orbits.

There is no perfect solution - higher altitude orbits may be better overall for naked-eye stargazers, but lower orbits are thought, based on the astronomy community's current understanding of satcon operation plans, to be better for large research facilities, which tend to be at lower latitudes.

\section{Discussion: Some Effects on Astronomy} \label{sec:disc}

With hundreds or even thousands of sunlit satellites moving across the night sky at a time, research astronomy will be affected.  
Other works discuss a broad range  of negative effects (such as pollution and collision risks) that satcons will have, for example, on the sustainable development of space \citep[e.g.][]{BoleyByers}, but here we focus specifically on the effects on research astronomy.
Following \citet{hainaut_williams2020}, we highlight in this section a few of the prominent negative effects, but this is by no means an exhaustive list.

\subsection{Bright streaks in research images and astrophotography} \label{sec:streaks}

The most obvious effect of satellites on optical research astronomy is the contamination of research images by  bright satellite trails, appearing as  lines cutting across the entire field of view of research or astrophotography images\footnote{Many astronomers have already experienced this, and it is documented widely.}. 
Research astronomers are developing algorithms to process these streaks, such as automatically identifying, characterizing, and removing them from images (see e.g., the SATCON2 Report). 
Such algorithms will help to mitigate some forms of data loss from science images and allow further image processing, but they cannot recover lost data.

If satellite trails are too bright, they can create cross-talk between CCD chips in large telescope, most well-documented for the LSST telescope \citep{TysonLSST}.
If satellite operators can reduce brightnesses below $g$ magnitude $\sim$7, then such cross-talks might be significantly reduced for some detectors, although this needs to be continuously evaluated. 
Further, if operators meet this brightness limit, it will fulfill the task of keeping most satellites below naked-eye visibility.  
SpaceX  has stated that this is the goal for their Starlink satellites -- a goal that has not yet been achieved \citep[e.g.][]{BoleySats,Mallama2021}.

The level of disruption to research astronomy will most strongly depend on satellite brightnesses and rates of motion. 
Consider Figure~\ref{fig:sunlit30}, which shows 400 satellites above 30$^\circ$ elevation for at least a portion of each night. 
This corresponds to a sky area of $\pi$ sr. 
The expectation that the satellite will pass through the FOV is then estimated as
\begin{equation}
    E = R w t N / \Omega_{\rm sky},
\end{equation}
where $R$ is the typical sky motion of a satellite, $w$ is the characteristic width of the detector (taken to be $1^\circ$, typical for a wide-field camera), $t$ is the exposure time, and $N$ is the number of satellites in the sky area $\Omega_{\rm sky}$. 
For $R\approx 2000$ $\arcsec$/s (typical for a satellite at $\sim$500km orbital altitude) and an exposure time of $t=10$ s, we find $E\approx 0.22$. Assuming Poisson statistics, the probability of one or more satellites passing through the FOV of the sq.~degree telescope is about 20\% under the models explored here.
For a 3 minute exposure, as is common for deep Kuiper Belt surveys \citep[e.g.,][]{ossos}, a typical exposure will include 4 satellite trails, depending on satellite brightness. 
While most parts of the world will have some period of time each night with few-to-no sunlit satellites, telescopes at high latitudes will experience this rate of satellite contamination all summer.

This calculation assumes that the satellites are independent of each other, when actually, their motions across the sky are highly correlated.  
Our rough calculations here agree with the much more thorough analysis of the number of satellites per telescope field and the fraction of lost exposures for different telescope facilities that is presented in \citet{Williams2021}.

\subsection{Severe effects on radio astronomy}

This paper is focused on optical light pollution, but we note that radio astronomy has the potential to be strongly affected by the abundance of transmitting satellites, particularly 
if the electronics that are used allow for out-of-band emission or overtones falling into bands otherwise protected for astronomy \citep[as happened with the Iridium satcon, see e.g.,][]{radiobook}.
Additional spectrum is further being sought after, and the use of frequencies that are typically used in terrestrial applications might be explored for on-orbit infrastructure, further restricting the frequencies available for scientific research.

The SATCON2 report, available to the public in October 2021\footnote{available at \url{https://doi.org/10.5281/zenodo.5608820}}, contains recommendations for how satcon operators can mitigate damage to radio astronomy.

\subsection{False positives in occultation studies}

Even when satellites are not sunlit, they can still potentially interfere with astronomical observations.  They are too fast to cause a measurable change in, for example, an exoplanet transit observation. But occultation experiments like TAOS, which is using high cadence observations with the goal of detecting occultations of stars by small bodies in the solar system \citep{Zhang2013}, could potentially experience a new source of noise.

A satellite occultation of a star lasts between about a tenth of to a full millisecond  \citep{hainaut_williams2020}, as  LEO satellite on-sky motions are of the order 2000~$\rm ''/s$, as noted above. 
While this timescale is much shorter than what is expected for an occultation by a small body in the outer solar system, an observing cadence of a few frames per second would cause this short dip to be confused with the $\sim$0.1~sec dip in brightness caused by a real occultation \citep{hilke2012,taosII}.  
With 65,000 satellites, satellite occultations could become frequent enough, depending on a telescope's field of view, to produce a false-positive signal as high as is expected for distant solar system small bodies.

To approximately calculate this, consider the same setup as in Section~\ref{sec:streaks}, with 400 satellites passing through a sky area of $\pi$ sr at any moment. 
At a range of 1000 km, a plausible angular size of a satellite is about $0.''5$. 
Taking the satellite's sky motion to be 2000 $''/\rm s$, then the satellite population would sweep out an area equivalent to that entire section of the sky with a frequency of about $0.3\rm~d^{-1}$. 
During that time, if we are monitoring an average of 10,000 stars \citep[as TAOS~II plans to do;][]{taosII} in each field, then for 8 hr nights, we would expect around 1000 satellite occultation events per night, on average.
This is much higher than the expected occultation frequency of stars by small trans-Neptunian objects (TNOs; $<10^{-3}$ per star per year), and may become a limiting source of noise for this type of astronomy research.

\subsection{Cumulative Effects: Total Sky Brightness}

As a result of adding more point sources, the total brightness of the night sky will increase.  For an estimate, we calculate a summed flux density using $V$ band magnitudes of all stars in the Hipparcos catalogue with $V<10$, converting to Janskys using the AB magnitude zero point, i.e., $m = -2.5 \log_{10}(f_\nu / 3631 \rm Jy)$. Averaging over the entire sky gives a surface brightness of 13.7 Jy/sq.~deg.  
We then repeat this calculation for our satellite models (using $V$ band estimates), including only the illuminated satellites and their corresponding brightness. For a latitude of 50$^{\circ}$ on the summer solstice at nautical twilight, satellite night sky brightness is about 0.5 Jy/sq.~deg., about 4\% of the starlight contribution.  

This simplistic calculation ignores any reflection by diffuse particles or by the atmosphere. 
Diffuse emission caused by sunlight reflecting off the huge number of tiny pieces of debris in orbit (``space junk'') has already been calculated to increase the overall brightness of the sky by 10\% above pre-industrial levels \citep{Kocifaj2021}.  
The placement of such large amounts of infrastructure in space raises the issue of the further generation of space debris, particularly if there is an on-orbit accident such as a collision or explosion, compounding the brightness problem.

\section{Conclusions} \label{sec:concl}

We acknowledge the broader engagement by SpaceX other satcon operators, such as OneWeb and Amazon (as in SATCON2), in entering discussions with astronomers concerning this issue. 
SpaceX has stated that they will engineer their Starlink satellites to be fainter than visual magnitude 7. We applaud their efforts,
but note that they have yet to achieve this goal.  
\citet{Mallama2021} presents observations showing that about 70\% of Visorsat Starlinks are brighter $V<7$ as observed from 39-44$^{\circ}$.
\citet{BoleySats} presents observations from 48.5$^{\circ}$~N, which show about 80\% of Starlinks are brighter than $g<7$, supporting the results of our modelling: the 550 km Starlink satellites are bright and visible throughout the night during the summer at moderate to high latitudes. 

Orbital altitudes of satellites are important, but there is no perfect strategy that will help all latitudes equally.  The current recommendation from the SATCON1 report is for satcon operators to use orbital altitudes of 600~km or less, and many of the operators have filed for orbits that primarily meet this requirement.
Such cooperation is good.
But as the modelling presented here shows, this altitude will not prevent satellites from being bright during the little but precious night available to  researchers and sky-watchers during summer at latitudes above approximately 45$^{\circ}$N and S.

The stated point of many of these new and future communications satcons is to bring high-speed internet to the world, benefiting many potential rural and remote users, especially isolated communities.
There are nonetheless important questions concerning the accessibility of such services \citep[e.g.,][]{Rawls2020}, particularly given the global impacts that bright orbital infrastructure could have, without mitigation. 
Such impacts are felt regardless of the degree of use or even access to the services, similar to classical pollution problems. 
Furthermore, orbital light pollution is inherently transboundary in that a state authorizing a satcon distributes that pollution worldwide, which has so far been done without broad consultation.
The simulations presented here show how the effects of satcons are a long-term change to the night sky worldwide, which is  a shared resource enjoyed and used by humanity \citep[see e.g.,][for further discussion]{CSAreport}.

There is simply no way to have tens of thousands of satellites in Low Earth Orbit and avoid  consequences for astronomy.  
But with strong, international cooperation and appropriate regulation of satellite numbers,  reflectivities, and satellite broadcasting, we can perhaps reach a compromise that allows much of research astronomy and traditional naked-eye stargazing to continue with minor losses.

\begin{acknowledgments}
The authors wish to acknowledge the land on which they live and carry out their research: Canadian Treaty 4 land, which is the territories of the n\^{e}hiyawak, Anih\v{s}in\={a}p\={e}k, Dakota, Lakota, and Nakoda, and the homeland of the M\'{e}tis/Michif Nation; the traditional, ancestral, and unceded territory of the Musqueam people, which has always been a place of learning for the Musqueam, who for millennia have passed on their culture, history, and traditions from one generation to the next on this site;  the traditional land of the Huron-Wendat, the Seneca, and the Mississaugas of the Credit, which today is still a meeting place and the home to many Indigenous people from across Turtle Island.  

This research has been supported in part by NSERC Discovery Grants RGPIN-2020-04111 (SML), RGPIN-2020-04635 (ACB) and RGPIN-2020-04513 (HR), as well as the Canada Research Chairs program.
\end{acknowledgments}

\software{This research was made possible by the open-source projects \texttt{REBOUND} \citep{ReinLiu2012}, \texttt{WHFast} \citep{ReinTamayo2015},
\texttt{Jupyter} \citep{jupyter}, \texttt{iPython} \citep{ipython}, and \texttt{matplotlib} \citep{matplotlib, matplotlib2}.}


\end{document}